# The effect of cities and distance on COVID-19 spreading in the United States

Troy McMahon[1], Shlomo Havlin[2], and Lazaros K. Gallos[1]

[1] DIMACS, Rutgers University, Piscataway, New Jersey 08854, USA.
[2] Department of Physics, Bar-Ilan University, Ramat Gan 52900, Israel.

**Abstract**

The COVID-19 pandemic has evolved over time through multiple spatial and temporal dynamics. The varying extent of interactions among different geographical areas can result to a complex pattern of spreading so that influences between these areas can be hard to discern. Here, we use cross-correlation analysis to detect synchronous evolution and potential inter-influences in the time evolution of new COVID-19 cases at the county level in the USA. Our analysis identified two main time periods with distinguishable features in the behavior of correlations. In the first phase, there were few strong correlations which only emerged between urban areas. In the second phase of the epidemic, strong correlations became widespread and there was a clear directionality of influence from urban to rural areas. In general, the effect of distance between two counties was much weaker than that of the counties population. Such analysis can provide possible clues on the evolution of the disease and may identify parts of the country where intervention may be more efficient in limiting the disease spread.

# INTRODUCTION

The COVID-19 virus has been spreading around the world since the beginning of 2020[1]. This pandemic can serve as a prototypical example of how highly infections viruses spread within our current modern, global society [2,3]. COVID-19 is an inherently infectious disease which is typically associated with localized outbreaks and infections due to close contacts [4]. In this pandemic, the global transmission rate and the geographical extent of the disease were greatly amplified compared to previously recorded pandemics [5,6], mainly because long distance transmissions are prevalent in a modern society, mainly enabled by large-scale global travel [7]. The rapid progress of the virus spreading is also the most closely monitored in history, which has enabled access to detailed records on the temporal and spatial evolution of the disease [8]. These high-resolution data allow a plethora of real-world analyses which have not been possible to this extent until now. More importantly, analysis of these data can provide crucial knowledge that can be useful to prevent or inhibit future pandemics.

Multiple works have studied the spreading patterns in various parts of the world and have considered a plethora of factors which may impact the virus spreading [9-14]. Many of these studies focus on regional effects that can impact the large-scale evolution of the disease [15-17]. Even though the virus transmission at a microscopic level can only occur at close contact between two individuals, the combination of population density and travel leads to a macroscopic expression of those close contacts [13]. The numerous local transmissions lead to outcomes that can be observed at the scale of a city, state, or country, leading to an emerging complex system where the details of person to person interactions are no longer important [18]. This consideration allows the study of spreading at a larger scale.

Here, we use a timeseries correlation analysis [19] of the COVID-19 epidemic data in the continental USA at a resolution of the county level. Our goal is to determine the degree of correlation in the disease evolution between different areas of the country, with a focus on the effect of distance between two areas and on the impact of population, such as whether spreading is similar in cities vs rural areas or if one area influence the other.

We find a rather weak effect of the distance between two areas. In general, neighboring areas tend to be highly correlated but the evolution of the disease in distant areas can also be significantly synchronized. We show that the reason behind this result is the effect of population. We detect a significant influence of the county population on our results so that urban areas tend to behave similarly even when separated by large distances. A time-shifted analysis [20] also reveals that there is an influence directed from urban areas towards rural areas. The evolution of the disease in rural areas seems to follow the evolution in rural areas by one or two weeks, and again this result does not depend strongly on the distance between the two areas.

We also find that both the intensity and the dependence of the correlations on the various factors have changed with time. During the first year of the epidemic, there were very few cross-correlated areas which were limited mostly to local areas. In the second year of the epidemic, correlations started increasing and becoming more widespread across the country. By the end of the second year, there was a significant degree of correlations in multiple areas of the country. These results may point out to the importance of the restriction measures which were in place for the majority of the first year of the epidemic but were loosened or removed in the second year.

**METHOD**

In this work, we use the New York Times dataset [21] which reports the daily new COVID cases in continental USA at the county level, based on reports from state and local health agencies. In our analysis, we aggregate the daily cases into a 7-day window to average out the daily results and provide a more consistent representation of the incidents in each county over time. As a result, our time unit is a week and we start our numbering with the week of February 1-Feburary 7, 2020, which we define as week $t_w=1$. Subsequent weeks are numbered accordingly, e.g. week $t_w=53$ corresponds to January 30-February 5, 2021. We then construct the timeseries of weekly cases in each county from week $t_w=1$ to week $t_w=132$, i.e. until the week of August 8-August 14, 2022.

We detrend the dataset by using the first difference method [19]. This means that we create a new timeseries for each county where each new point in a given week $t_w$ corresponds to the difference of the value at this point and the value at the previous time step $t_w$-1 in the original data. These detrended series are used for all the calculations in this study.

In our analysis, we only keep counties with a population larger than 50,000 people. This leaves us with 978 counties out of a total of 3108. We calculate the cross-correlations between the detrended timeseries corresponding to any two counties in this set by using a window of 20 weeks, i.e. the correlation calculation is based on series which include a total of 20 values centered in a given week $t_w$ (the results remain similar when we increase the window to 30 or 40 weeks). We then use a moving time window and we can calculate how the correlation changes between any two counties as time progresses.

We employ a time-lagged cross correlation method to detect possible directionality links, where the timeseries of one county is shifted by an offset (lag) of $s$=-5 to $s$=+5 weeks. We calculate correlations through the Pearson coefficient. We use the following formula for two differenced timeseries $x_1(t)$ and $x_2(t)$ at week $t_w$ for lag $s$:

$$C(t_w, s) = \frac{\sum_{t=t_w-10}^{t=t_w+9}[x_1(t) - \overline{x_1(t_w)}][x_2(t+s) - \overline{x_2(t_w)}]}{\sqrt{\sum_{t=t_w-10}^{t=t_w+9}[x_1(t) - \overline{x_1(t_w)}]^2 \sum_{t=t_w-10}^{t=t_w+9}[x_2(t+s) - \overline{x_2(t_w)}]^2}}$$

The average values $\overline{x_1(t_w)}$ and $\overline{x_2(t_w)}$ are calculated over the same time period as the limits in the sum, e.g. $\overline{x_1(t_w)} = (\frac{1}{20})\sum_{t=t_w-10}^{t=t_w+9} x_1(t)$. The value of $C(t_w,s)$ is a measure of the 'synchrony' or a link between two timeseries in different countis, i.e. how correlated is the evolution of new COVID cases between two counties. If the peak correlation occurs at $s$=0 then the disease follows similar patterns concurrently in both counties and it is probable that external factors influence the evolution paths or that the disease naturally spreads in similar ways. If lagged correlations are significantly stronger than unlagged ones, then this is an indication that one county may lead, or somehow influence, the evolution of spreading in the other county over a time interval equal to the lag value of the peak.

# RESULTS

## Unlagged correlations

Our first goal is to detect statistically meaningful correlations between different areas in the country and identify whether we can find and explain the similar disease evolution patterns according to certain salient features. Such analysis can also help with the evaluation of quarantine measures. Effective measures would lead to a breakup of correlations and ideally only allow random independent variations in each county. On the contrary, if strong correlations emerge in large areas of the country, then this can be regarded as an indication that the disease is largely synchronized and restrictive measures did not inhibit the outbreak. In this section we restrict the analysis to the case of identifying correlations with no lag, i.e. $s=0$ in Eq. (1).

For each week between $t_w=10$ and $t_w=120$ we calculate correlations between all possible pairs of the 978 counties. We then impose a threshold of $C(t_w)>0.9$, so that we only consider the strongest correlations. Fig. 1a shows a map of these correlations at different times with connections between counties whose correlation exceeds the 0.9 threshold. The number of strong connections increases with time until roughly the middle of 2021, when this number starts to drop. By the end of 2021 a second wave emerges and the number of links becomes higher again continuing at an increasing pace into 2022.

We can consider this representation as a network where the nodes (counties) are connected through the strong correlations. We can isolate the largest connected cluster in the network, which is a measure of the extent of the country where COVID spreading was progressing in similar ways. The maps in Fig. 1b indicate that the largest cluster remained largely localized until roughly the end of 2021 while a spanning cluster emerged at the later stages of the pandemic, and this cluster includes almost all the counties in the United States.

The time evolution of the largest cluster size, $S_{max}$, in Fig. 1c provides a detailed picture on how this size changed with time. We can see that there are four main peaks. The first two peaks on September 2020 and March 2021 are relatively small with $S_{max}$ reaching roughly 30% of the country. The peak around June 2021 is much stronger at around 55%, while from November 2021 until April 2022 the spanning cluster includes almost all counties. It is interesting to note that the first three peaks emerge at times when COVID spreading was in decline, as can be seen by the plot of new COVID cases. The fourth peak also starts when new cases are low, but the value of $S_{max}$ remains large during the outbreak at the end of 2021.

These results indicate a change in the evolution of epidemic spreading during this time interval. Up until the autumn of 2021, when many restriction measures were still in place, correlations were increasing when the number of new cases was declining and a relatively small fraction of the counties participated in the largest cluster. The outbreak in December 2021 is qualitatively different because the largest cluster spans practically the entire country and it remains intact throughout the peak. Since correlations quantify the degree of 'synchronized' evolution, the results here suggest that at the second phase of the epidemic, starting roughly in September 2021, many parts of the country were following similar patterns in the number of new cases.

It is probable that there is an underlying reason to explain why certain county pairs exhibit strong correlations while others do not. One obvious answer is proximity. There is always a higher probability that two neighboring counties will follow similar changes in spreading compared to distant counties. To study this effect, we show in Fig. 1d the time evolution of the average correlation as a function of the distance between the geographical centers of two counties. As expected, strong correlations are observed at short distances at all times. When the average correlation increases significantly over the entire country, as shown by the peaks in Fig. 1c, then correlations are largely independent of the distance. Otherwise, the correlations depend only weakly on the distance, with the exception of very small and (some times) very large distances that span the entire country. There are some differences in weeks 20-50, when intermediate distances are less correlated and the same is true in weeks 70-80. These small variations indicate that distance may not be the main factor that drives correlations, so we proceed to study the effect of population.

We start by identifying 'persistent' links, i.e. pairs of counties whose correlation remains consistently strong over time. For each pair of counties, we calculate the average correlation from $t_w$=20 to $t_w$=110 and we define as persistent a link where this average exceeds a value of 0.7. A map of these links is shown in Fig. 2a, where we separate local links (i.e. counties within the same state) from long-range links (i.e. counties at different states). It is quite clear in the map that long-range links predominantly connect large urban centers, located mainly in the Northeast, Midwest, Florida, and West Coast. In Fig. 2b we calculate the distribution of population in the counties which are connected through these persistent links. We bin the population distribution of all 978 counties into 10 boxes, so that each box contains roughly a $10^{th}$ percentile of the distribution. This corresponds to 98 counties in each box, with the exception of the last box which contains 96 counties. If the persistent links were randomly distributed in terms of population we would expect a uniform distribution at around 10%. The results in the plot indicate a largely skewed distribution where high population counties (largely corresponding to urban areas) are over-represented and low-population counties (rural areas) are under-represented. Almost half of the counties in long-range persistent links are in the highest population box, which includes populations between 670,000 and 10 million people. Within the long-range persistent links, there are very few counties with population less than 125,000 people, which is roughly the median value in the distribution. Even the local in-state links follow a similar pattern where counties with large populations appear much more frequently than expected by a random distribution of the persistent links, while small population counties are still under-represented.

This is already a strong indicator that correlations may be largely influenced by the size of the population, which would correspond to an urban city vs a rural area. This is probably due to the fact that large cities have higher probability of mutual visits which increase the probability of synchronization in the epidemic. To further test this hypothesis, we calculate the average correlation between counties of population in the different bins described above. In Figure 3, we present the values of these correlations in a matrix form at different times. Since there is no directionality in these calculations, the matrix is symmetric along the diagonal. It is clear that the correlation at the upper right part of the matrix (which corresponds to both counties having large populations) is significantly stronger compared to e.g. the bottom left part which corresponds to both counties having low populations. The correlations between high and low populations have in general intermediate values, as can be seen by the columns at the right part of the matrix (or the top rows). After 90 weeks there is a global increase of correlation and the population does not seem to have a significant impact any more, since the majority of the correlations already have large values.

The above analysis provides significant support to the hypothesis that population plays a much larger role in synchronized correlations, especially compared to the effect of the distance.

**Lagged correlations**

The results in the previous section were based on the correlation between two timeseries evaluated at the same time. In this section, we consider the case of time lagged correlations, where the value of lag, $s$, in Eq. (1) is no longer zero. We studied lags in the range of $s=-5$ to $s=5$ weeks, but there were no significant correlations when the lag exceeded 3 weeks.

In Fig. 4 we present the time evolution of the correlation matrix at different lags. The matrix displays the average correlation between counties, which are grouped according to their population. For comparison, we also include the unlagged case of $s=0$, which was described above. One key result is that for the first year of the epidemic, roughly until March 2021, we did not observe any significant lagged correlations at any lag value except for $s=0$, so we do not show these cases in the figure. This result suggests the absence of direct time-delayed influence from almost any geographical area to any other. Except for the 20-week interval beginning March 2021, lagged correlations remain weak until the interval starting at July 2021. It is interesting to see in the panels after that time, that the majority of strong correlations for positive lag values appears predominantly at the right part of the matrix, i.e. high population counties leading (earlier in time) lower populations by one or two weeks. This strong directionality is also seen at negative lags, where the majority of correlations is located at the top part of the matrix, i.e. low-population counties follow high population counties. There are a few negative correlations that emerge at a lag of three weeks, but in general they are weaker than those for lags of $s=1$ or $s=2$. In summary, the panels in Fig. 4 show that influence is strongly directed from high population to low population areas.

This result is verified through the time evolution of the lagged correlations at the four corners of the matrix, which correspond to counties of high-high population, high-low, low-high, and low-low (where the first category precedes the second). Fig. 5 shows that these curves in general follow the average time evolution of the correlations and they become weaker with increasing lag. However, it is important that for lag $s=1$ and $s=2$, the strongest lagged correlations appear in the case of high-high populations, but at the later part of the epidemic the high-low case becomes the strongest. On the contrary, the low-high case is the weakest one with values close to 0. This result is mirrored in negative lags where the low-high case shows the strongest correlations, and essentially delivers the same message that the timeseries of new cases in high population counties precede those of low population.

For a direct support of the links directionality, we calculate the average number of connections that originate in counties with high population and compare with those originating in counties of low population. Similarly, we calculate the average number of incoming connections in each case. Figure 6 displays these results as a function of the population at different times. Starting from lag $s=0$ we can see that high population counties have a higher number of both incoming and outgoing connections compared to counties of lower population, a result which is consistent with our findings in the previous section. For all positive lags, the number of outgoing links from urban areas is significantly larger than those in rural areas, which in general remain close to 0. The opposite picture is found for negative lags, which of course carries the same message of rural areas following the urban areas. The low population counties are over-represented in the number of incoming links in most cases for positive lags, but the curves are not as clearly monotonic and in many cases there is a largely uniform distribution across all population sizes. The general result points to the idea that the source of time-shifted correlations are high-population counties which drive the epidemic of mostly low-population counties with a lag of 1 or 2 weeks.

Finally, we study the combined effect of distance and population in time-lagged correlations. In Fig. 7 we compare the average correlation between urban-urban, urban-rural, rural-urban, and rural-rural areas as a function of the distance. The case of no lag, $s=0$, is consistent with the results reported above, where there is a small dependence on distance for the global average, with urban-urban correlations always remaining above the average and rural-rural correlations being below this average. For non-zero values of the lag, there is almost no distance dependence and the correlations do not change much with the distance. At the first phase of the epidemic, e.g. January-June 2021 in the plot, we see larger than average correlations between urban areas and lower than average between rural areas. In late 2021 and early 2022, when correlations throughout the country have risen significantly, the urban to rural influence are above the average correlation at all distances for positive lag values, while the opposite rural-to-urban influence is much smaller. This result provides additional support that the epidemic has been largely driven by large urban centers which can influence many areas around the country, almost independently of the distance and population.

## DISCUSSION

The analysis of the timeseries cross-correlations between different locations of new COVID-19 cases in the USA can provide important information on the factors which influence the long-range virus spreading. In this work, we analyzed how cross-correlations changed with distance between two counties and with population of the counties. Based on the behavior of these cross-correlations, we can roughly classify the first 2.5 years of the epidemic in two phases: the first phase covers the time period between early 2020 and mid-2021, and the second phase is between mid-2021 to mid-2022.

The first phase is characterized by generally weak correlations and by the absence of directed influence, as seen by the small values of time-lagged correlations. During this phase, the only significant correlations have been observed between areas of high population at no lag, and they remain strong almost independently of the distance between these urban centers. Why is the weekly number of new cases in cities similar throughout the country and different than those in rural areas? A plausible explanation is that there is a large daily volume of traffic between cities so that infection rates can be indirectly synchronized through travel, but any such influence would be below the resolution of one week used in our study. Another explanation is that the disease patterns may be influenced by global features of urban areas, which drive the epidemic in similar paths even in the absence of direct influence between two cities. These factors could be related to static urban features, such as dense population, easier access to health care, etc, or due to dynamic processes related to urban lifestyle, e.g. public transportation or increased daily activities leading to increased interactions with many people [9, 22, 23]. Any correlations in this phase, other than the synchronized inter-city correlations, are very small independently of the amount of lag. This can be seen as a desirable outcome in terms of public health because a vanishing correlation means that the disease follows independent patterns in different areas with the implication that the extent of influence remains small and local outbreaks cannot propagate far. Interestingly, quarantine and travel restriction measures were implemented and remained in force for the most part of this time period. It is possible then that the lack of correlations that we observed can be attributed, at least to some extent, to these measures.

In the second phase of the epidemic, we observed a significant increase in the strength of cross-correlations throughout the entire country. Additionally, the time-shifted correlations indicate a strong directionality of the correlations from large urban areas to rural areas with an offset of one to two weeks. The combination of these two factors in this period coincides with the relaxation and complete elimination of restriction measures, as well as with the presence of the largest peak in the number of global cases in the country, which occurred in January 2022. We should note that strong correlations do not necessarily imply extended spreading and they are not dependent on the absolute number of cases in the country. In practice, cross-correlation among different counties implies some kind of synchronous time evolution, or some directionality of influence, even if the number of cases is very low. For this second phase, the main outcome is that the evolution of the disease seems to become synchronized over large areas of the country with the potential to influence each other.

## ACKNOWLEDGEMENTS

This work was supported by a joint NSF-BSF grant. TM and LKG were supported by NSF through DEB-2035297. SH was supported by BSF through grant 2020645.

**(a)**

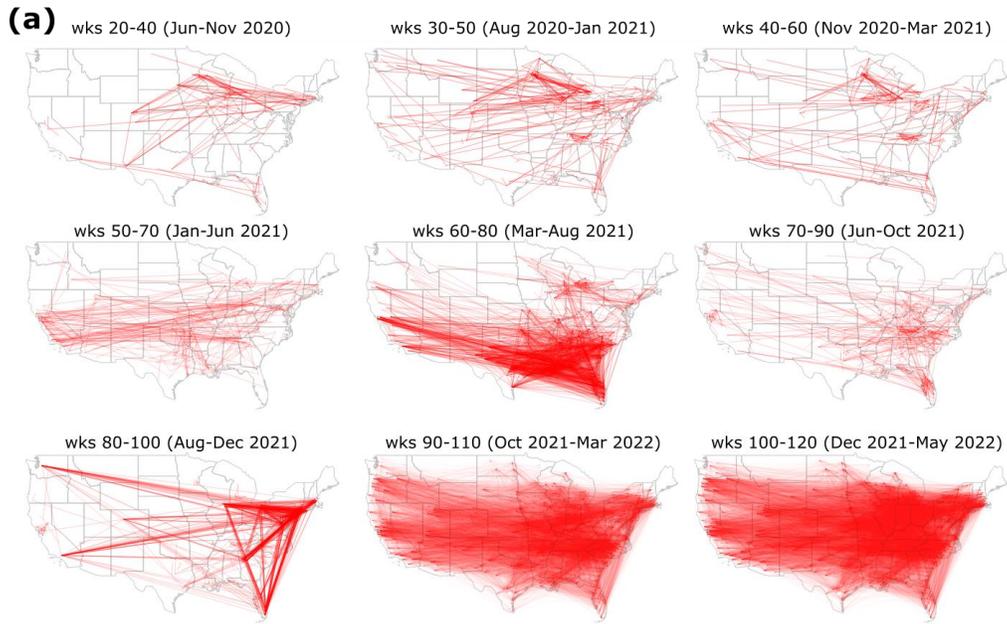

**(b)**

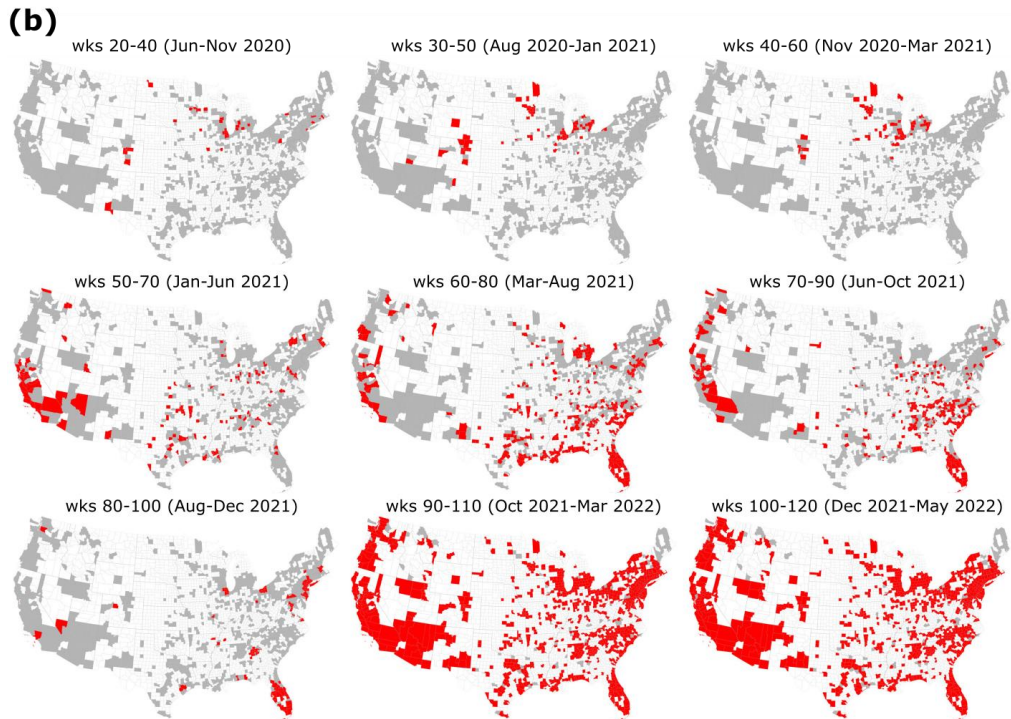

**(c)**

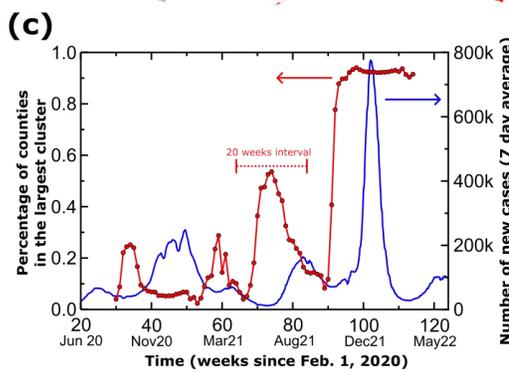

**(d)**

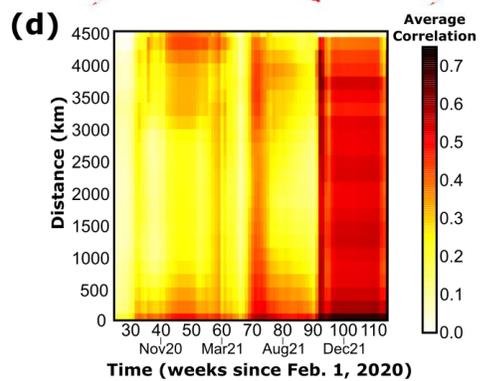

**FIGURE 1 | Clusters of high correlation.** (a) The links connect all counties whose correlation is above 0.9 for the 20-weeks time interval shown in the figure. (b) Counties participating in the largest connected cluster are shown in red. We only consider counties with population larger than 50000, which are shown in gray. (c) Time evolution of the largest cluster size. The fraction of counties participating in the largest connected cluster (red line), as determined by correlations larger than 0.9, is compared to the mean number of new cases in the country (blue line). Each point in the largest cluster size curve corresponds to the correlation calculated within a time window of 20 weeks (shown in dotted line) and the point is centered within this interval, e.g. the size of the cluster at $t_w$=70 was based on correlations in weeks 60-80. (d) Effect of the distance on correlation. The vertical axis shows the average correlation between any two counties in the given distance at different times.

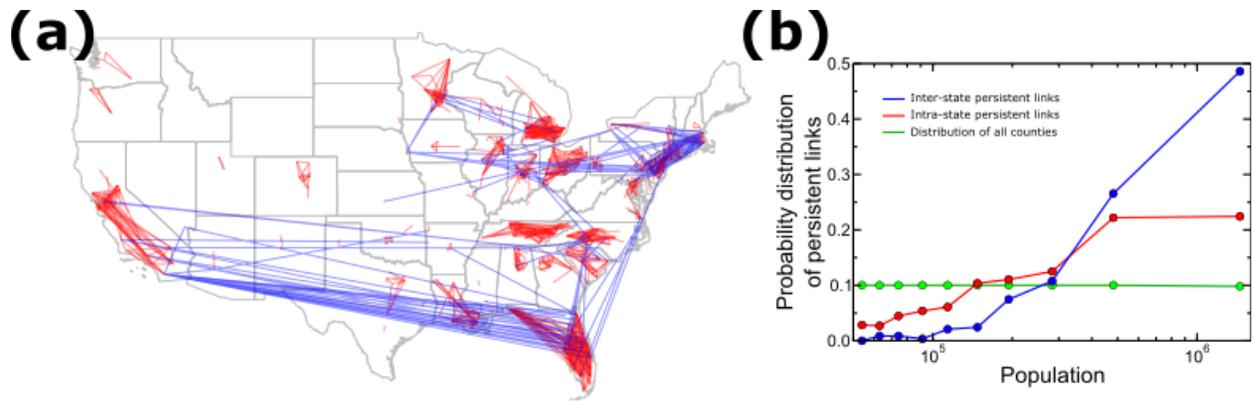

**FIGURE 2 | Persistent links (a)** The map shows the 'persistent' links, i.e. those whose average correlation over all times is higher than 0.7. Red links indicate "local" links, i.e. those within the same state, and blue links connect 'remote' counties, i.e. those at different states. **(b)** The population distribution of counties at the edges of persistent links. The population distribution of counties within the same state is shown in red and the distribution across states is shown in blue. For reference, the distribution of population of all counties of population greater than 50,000 is shown in green.

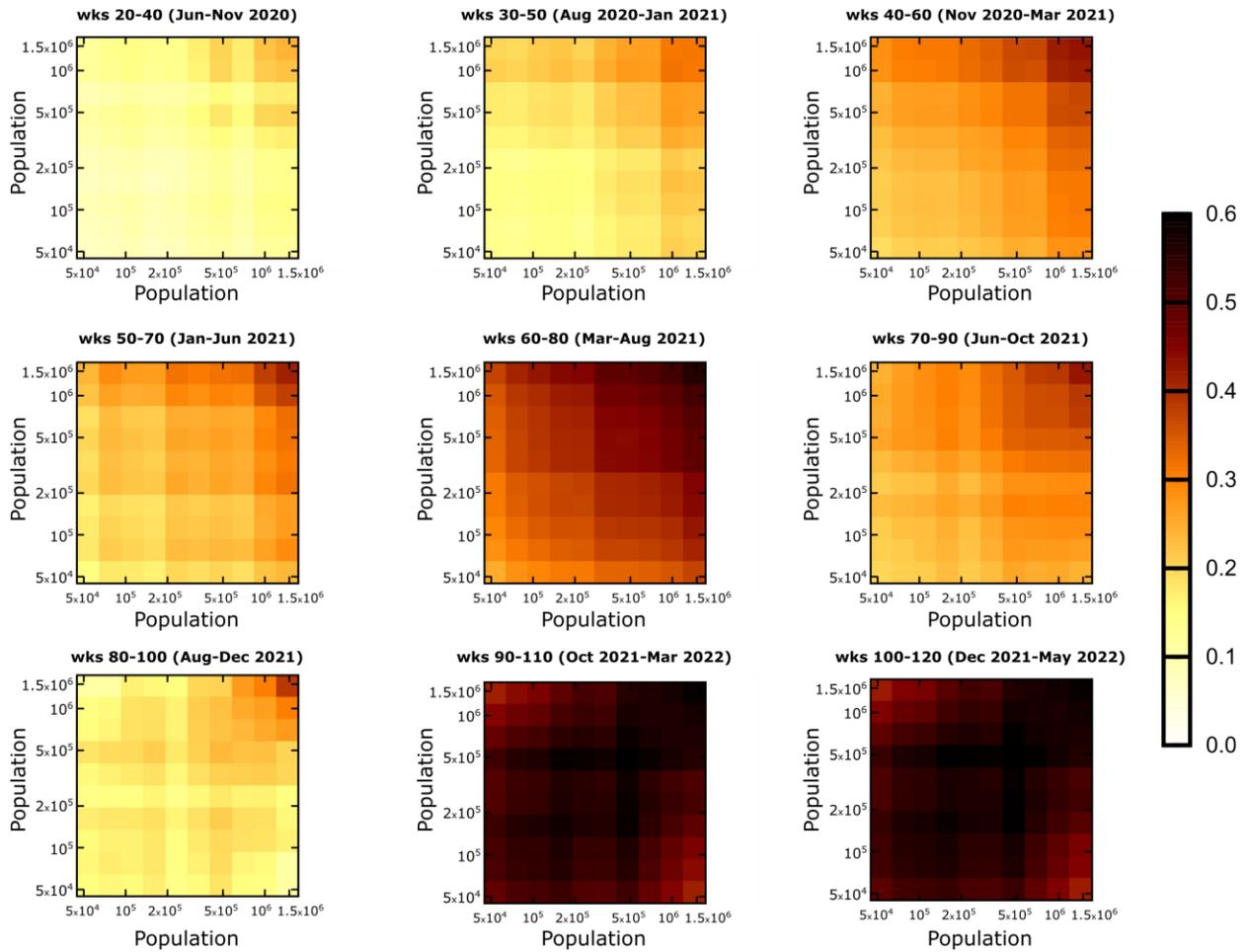

**FIGURE 3 | Effect of the population on correlations.** The colors in each plot indicate the average correlation between counties of given populations at different times. Correlations between counties of high population (top right) are always stronger than between combinations which include low population counties. Starting at the end of 2021, correlations become much stronger in intensity almost uniformly and independently of the population.

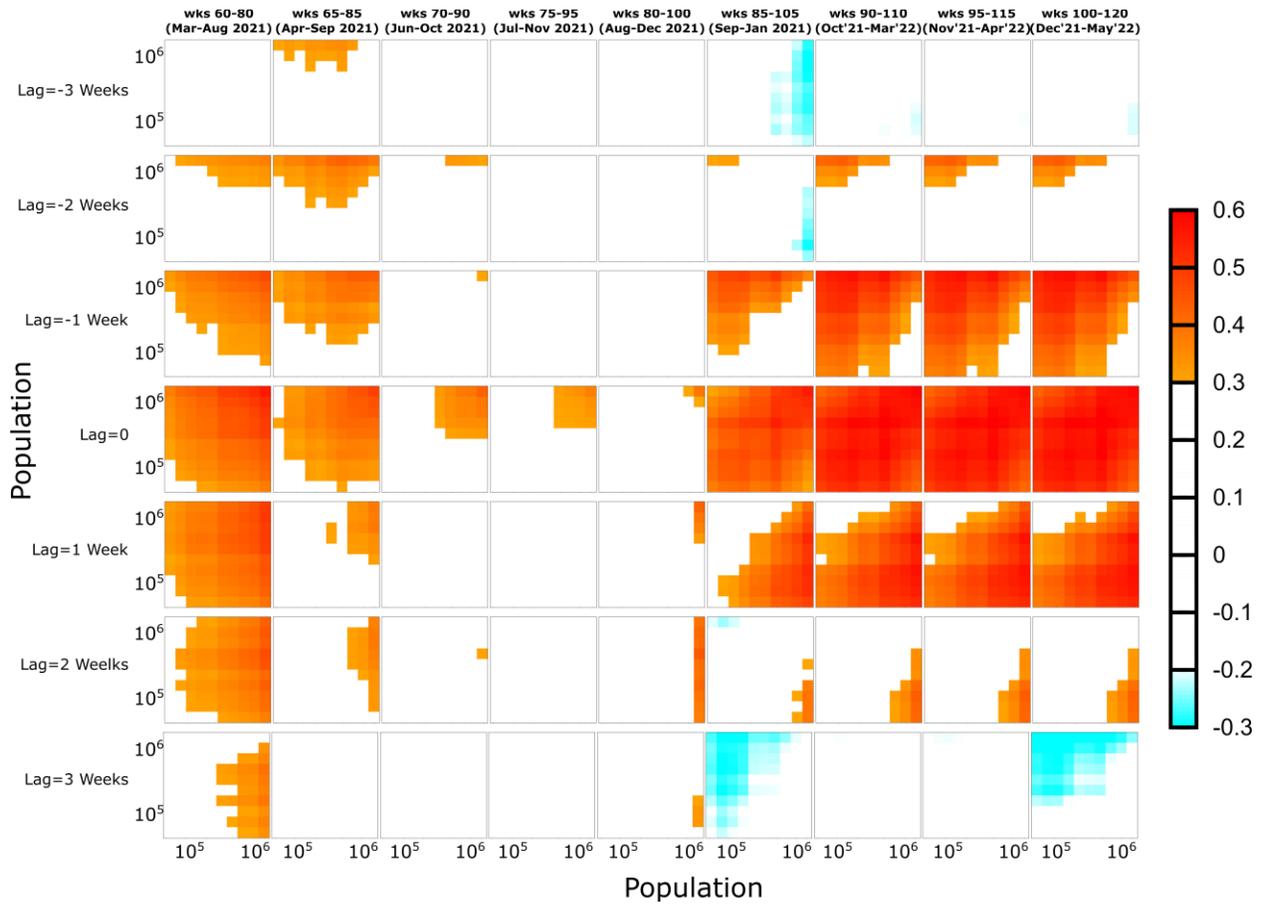

**FIGURE 4 | Lagged correlations indicate that high-population counties may influence low-population counties.** Each row corresponds to a different value of the lag from *s*=-3 (top) to *s*=+3 (bottom) (the row for *s*=0 shows the same data as Figure 3). Consecutive plots in a row represent the time evolution of average correlation between counties of varying population at the given lag value. For clarity, we only show values of correlation which are higher than 0.3 or lower than -0.2. There were no lagged correlations higher than 0.3 until March 2021, except for *s*=0 as shown in Fig. 3.

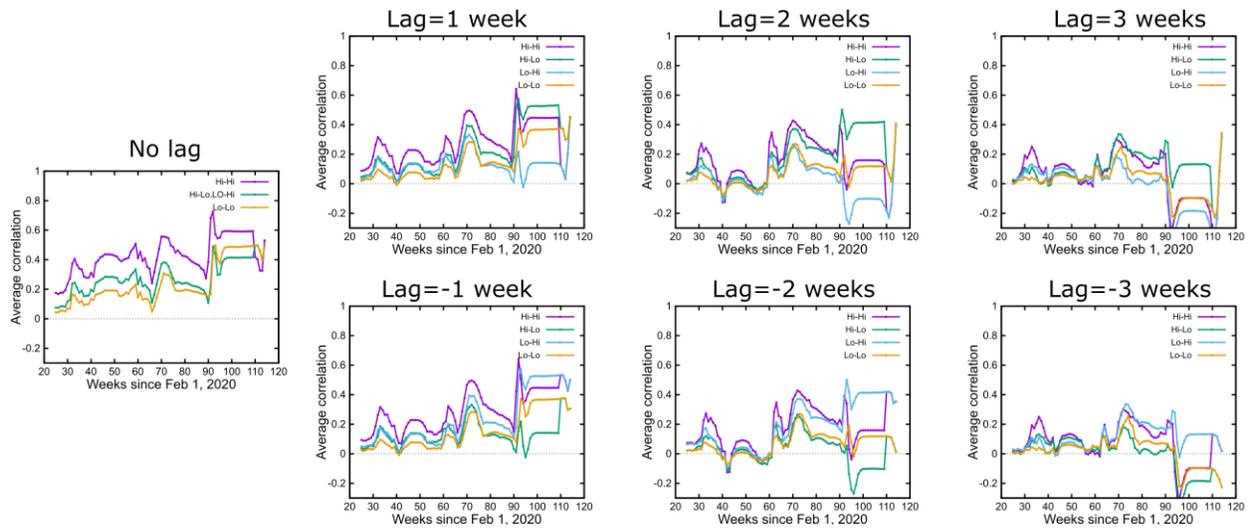

**FIGURE 5 | Lagged correlations between counties of high and low populations as a function of time. (a)** Each plot corresponds to a different value of lag. The lines show how the average correlation changes as a function of time between the combination of counties in the highest- and lowest-population bins. Notice that correlations between high-population and positively lagged low-population counties (green lines) are always stronger than the reverse low-to-high population (blue lines). The opposite is true for negative lags.

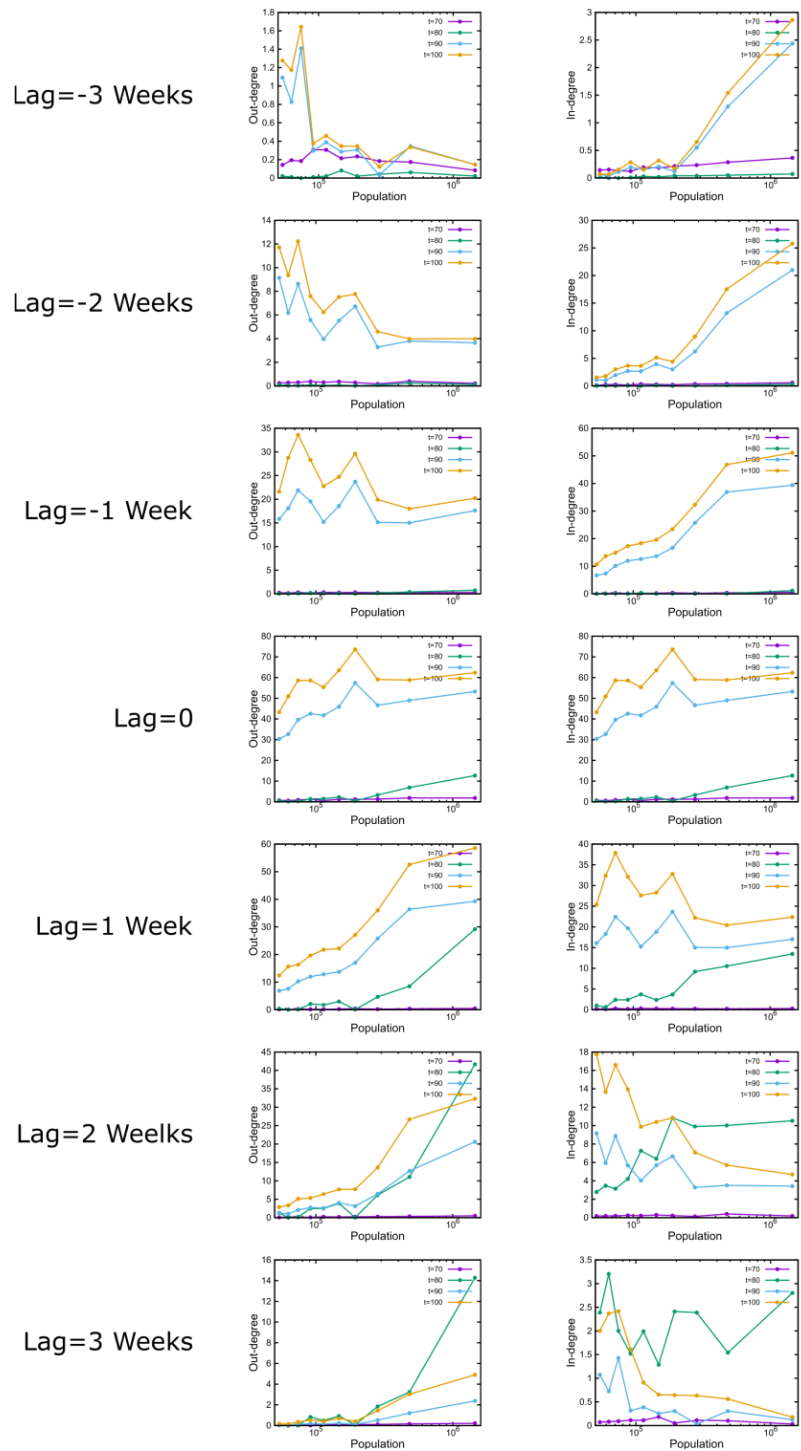

**FIGURE 6 | The source of lagged correlations are predominantly the high-population counties.** The plots on the left column show the average number of counties in a given population bin which are at the source of lagged correlations with a value higher than 0.9. On the right column, the lines show the average number of counties in a given population bin at the receiving end of the correlation. From top to bottom, the lag increases from *s*=-3 weeks to *s*=+3 weeks.

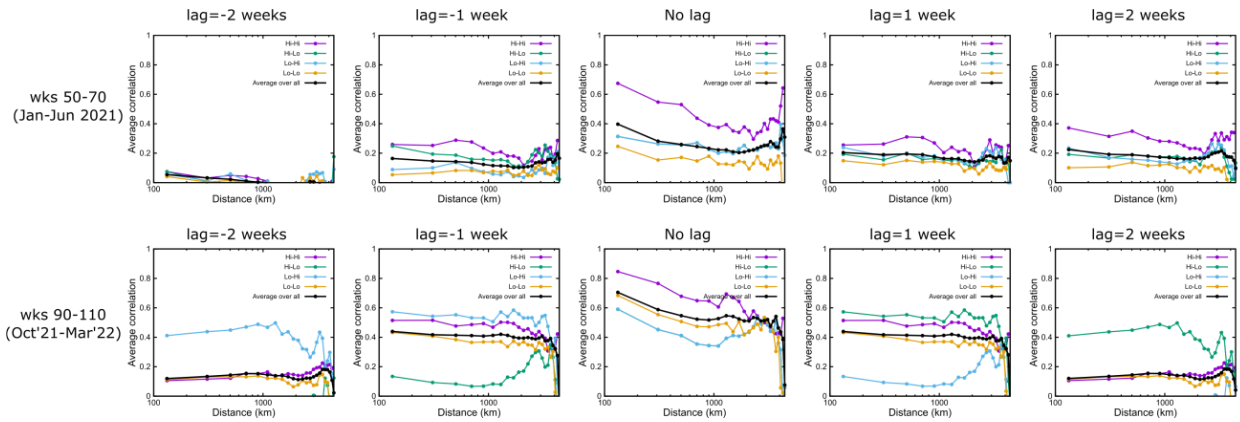

**FIGURE 7 | Correlation as a function of distance for counties of different population at different lag values.** The average correlation as a function of distance between counties of high and low population. The black line indicates the average over all counties at the given distance independently of population. The panels in each column display the results for lag values ranging from $s=-2$ to $s=+2$, as shown in the figure. The top row corresponds to the interval January-June 2021 and the bottom row to the interval October 2021-March 2022.